# Mechanisms of particle entrainment in confined gas–particle systems under moving boundaries

Arata Hashimoto[1], Ryosuke Mitani[1], Toshiki Imatani[1], Mikio Sakai[1*]

*1 Department of Nuclear engineering and management, School of Engineering, The University of Tokyo, 7-3-1 Hongo, Bunkyo-ku, Tokyo 113-8656, Japan*

Corresponding authors:

* mikio_sakai@n.t.u-tokyo.ac.jp (Mikio Sakai)

**Abstract**

Particle entrainment in confined gas–particle systems driven by moving boundaries plays a central role in a wide range of industrial and natural processes, including pharmaceutical tablet manufacturing, food processing, and chemical engineering operations. Such transport is often loosely referred to as a "suction effect," yet its physical origin and governing principles have remained unclear, particularly in the presence of highly unsteady flow, strong particle–particle interactions, and complex transient force networks. Here we investigate suction-induced particle entrainment in a prototypical confined system using high-fidelity coupled CFD–DEM simulations that resolve the unsteady gas flow and track the discrete particle dynamics under moving boundary conditions. Using an advanced moving-boundary CFD–DEM framework, we decompose the forces acting on individual particles and quantify their temporal evolution. We show that suction is not a purely pressure-driven phenomenon, but arises from the combined action of pressure-gradient and unsteady drag forces generated by boundary-accelerated gas flow. Despite the strongly heterogeneous and transient nature of the flow and force fields, we demonstrate that the final entrained mass is governed to leading order by a single energetic measure, namely the mechanical work performed on the particle assembly along the entrainment direction over the duration of boundary motion, rather than by the peak magnitude of instantaneous forces. Systematic variation of boundary kinematics reveals that extending the boundary displacement or modifying its boundary-velocity history regulates entrainment primarily by altering the time over which fluid–particle forces perform work, even when their instantaneous intensities differ substantially. These findings uncover a previously hidden organizing principle in suction-driven entrainment and establish a work-based physical framework for boundary-induced particle transport in confined gas–particle systems, providing a unified interpretation applicable across a broad range of engineering and natural environments.

**Introduction**

Particle transport in confined fluid–particle systems (1–11) is a fundamental process in a wide range of natural phenomena and industrial technologies. Examples include pharmaceutical tablet (12, 13) and capsule manufacturing (14), food processing (15, 16), and engineering operations such as catalyst production (17) and additive manufacturing (18–20), as well as geophysical and biological flows (21–25) in which particulate matter is transported within bounded geometries. In these systems, the interaction between the dispersed particles and the surrounding fluid, together with geometric confinement by solid boundaries, governs mixing, filling, and material uptake, and thereby plays a central role in determining process efficiency and functional performance.

Among the various mechanisms that drive particle transport in such confined systems, the motion of solid boundaries constitutes a particularly important and ubiquitous class. When objects such as a wall, piston, membrane, or deforming surface move, they generate unsteady flow in the surrounding fluid, which in turn entrains particles into regions that would otherwise remain poorly filled or weakly mixed. This boundary-driven particle transport is commonly described in terms of a "suction effect," whereby particles are drawn into a cavity or void as the boundary recedes.

From a fluid–particle interaction perspective, confined systems with moving boundaries represent a class of unsteady multiphase flows in which momentum is transferred from the boundary to the fluid and subsequently to the dispersed phase. In such systems, the instantaneous magnitude of the fluid-induced forces alone is insufficient to characterize particle transport. Instead, particle motion and uptake are determined by the cumulative mechanical work performed by these forces over the duration of boundary motion. Despite its fundamental importance, this work-based interpretation of suction-induced entrainment has received relatively little attention, largely because direct experimental access to transient pressure fields, drag forces, and their time integration at the particle scale is extremely challenging. Moreover, in confined gas–particle systems with dynamically moving boundaries, the flow field inside the newly created cavity is highly unsteady and spatially complex, involving rapid acceleration, deceleration, and recirculation of the gas phase as well as strong particle–particle interactions. Under such conditions, it is far from obvious a priori that particle entrainment and filling can be characterized by a single mechanical measure such as the accumulated work, rather than by more detailed features of the instantaneous pressure or velocity fields. This intrinsic complexity has hindered the formulation of a simple and universal physical description of suction-driven entrainment in moving-boundary multiphase flows.

High-fidelity numerical simulations that resolve both the unsteady gas flow and the discrete particle dynamics offer a powerful means to overcome these limitations. In particular, coupled computational fluid dynamics–discrete element method (CFD–DEM) approaches (26–29) equipped with accurate moving-boundary representations (30, 31) make it possible to decompose the forces acting on individual particles and to quantify their temporal evolution and accumulated work. Such methods enable a direct assessment of how pressure-gradient and drag forces arise from boundary-driven gas acceleration and how they jointly control particle entrainment in confined geometries.

In this study, we employ a coupled CFD–DEM framework (29–34) with an advanced moving-boundary model to investigate suction-induced particle entrainment in a prototypical confined gas–particle system. By explicitly resolving the unsteady flow generated by boundary motion and the resulting fluid–particle and particle–particle interactions, we quantitatively decompose the total force acting on the particles into pressure-gradient, drag, and contact contributions and evaluate the mechanical work performed by these forces. We show that the suction effect does not originate from a pressure deficit alone, but from the combined and time-integrated action of unsteady drag, pressure-gradient forces, and interparticle contacts transmitted through the dense granular assembly. Furthermore, we demonstrate that the efficiency of entrainment and overfilling is governed by the total mechanical work imparted to the particles over the duration of boundary motion, rather than by the instantaneous magnitude of any single force component, and that increasing the excursion of the moving boundary enhances entrainment primarily by extending the period over which this work is accumulated.

The physical mechanisms identified here provide a unified work-based interpretation of suction-like entrainment in confined gas–particle systems under moving boundaries. This framework clarifies why descriptions based solely on transient pressure fields, as commonly employed in previous experimental and theoretical studies, are insufficient to capture the physical origin of suction-driven transport, because they neglect the essential roles of unsteady drag, particle contacts, and their accumulated mechanical effect. Although the present analysis is illustrated using a representative industrial configuration, the force- and work-based perspective developed in this study is expected to be broadly applicable to boundary-driven particulate transport in pharmaceutical, food, chemical, and additive-manufacturing processes, as well as in geophysical and biological systems where particles are entrained by unsteady flows in confined geometries.

**Results and Discussion**

**Physical mechanism of suction-driven entrainment.**

In confined gas–particle systems with moving boundaries, the flow inside the newly formed cavity is highly unsteady and spatially inhomogeneous, involving rapid gas acceleration, strong particle–particle interactions, and complex transient force networks. Under such conditions, it is not obvious a priori that the final amount of entrained particles can be described by a single energetic quantity, such as the accumulated mechanical work, rather than by detailed features of the instantaneous pressure and force fields. We therefore first examine suction-induced particle entrainment under two boundary kinematics with identical final displacement and total descent time, but different histories of boundary velocity. In both cases, particles are initially located in an upper compartment, and a rigid boundary element forming part of the bottom wall retreats downward, creating a new void region beneath the particle bed. The resulting unsteady gas flow entrains particles into this newly formed cavity. By prescribing either a single-stage constant-velocity retreat (Case A1) or a three-stage piecewise constant-velocity retreat with increasing speeds (Case A2), while keeping the total displacement and duration identical, the overall extent and time scale of boundary motion are fixed and only the temporal evolution of the boundary velocity is varied.

Figures 1A and 1B show snapshots of the boundary motion and particle distributions for the single-stage constant-velocity case (Case A1) and the three-stage piecewise constant-velocity case (Case A2), respectively, at the same elapsed times. Although both motions reach the same final position within the same duration, their transient entrainment dynamics differ markedly. These differences are quantified in Fig. 1C, which presents the time histories of the particle mass flux into the cavity and the accumulated entrained mass. In Case A1, the mass flux remains nearly constant throughout the downward motion of the boundary and decreases rapidly once the boundary reaches its final position. In Case A2, by contrast, the mass flux increases stepwise in accordance with the successive increases in the boundary velocity, reaches a higher maximum, and then decreases after the boundary stops. Thus, the three-stage motion generates a stronger but shorter-lived suction response, whereas the single-stage constant-velocity motion produces a weaker but more sustained entrainment. Despite these pronounced differences in the instantaneous mass flux, the time-integrated flux, namely the total mass entrained into the cavity, is nearly the same in the two cases. During the early stage of the motion, particle filling proceeds more rapidly in Case A1, whereas in Case A2 entrainment is initially weaker but becomes stronger at later times as the boundary velocity increases in steps. As a result, the integrated filling curves converge, leading to almost identical final filling amounts. These observations demonstrate that the final amount of entrained mass is not determined by the peak strength of the suction-induced flow, but by the accumulated entrainment over the entire duration of boundary motion. This behavior suggests that descriptions based solely on instantaneous pressure deficits or peak force levels are insufficient, and that an energetic measure accounting for both force magnitude and duration is required to capture the governing physics of suction-driven entrainment.

To further clarify the physical origin of this behavior, we analyze the mechanical power and work performed on the particles by the total force acting on them, which includes the pressure-gradient force, the drag force, gravitational force, and contact forces arising from particle–particle and particle–wall interactions. The instantaneous power is defined from the components of the particle velocity and the total force along the entrainment direction, whose time integral gives the mechanical work delivered to the particle assembly along that direction. Figure 2A shows representative snapshots of the pressure-gradient and drag forces acting on the particles during the suction process for Cases A1 and A2. In Case A1, both forces exhibit large magnitudes over a very short time interval immediately after the onset of boundary motion. In Case A2, similarly strong force responses appear during the short periods when the boundary velocity changes stepwise. In both cases, the pressure-gradient and drag forces are significantly larger than gravity, indicating that they constitute the dominant driving forces for particle entrainment during suction. Figure 2B presents the temporal evolution of these two force components projected onto the entrainment direction. Throughout the period of active suction, the pressure-gradient force is consistently larger than the drag force; however, the difference in magnitude is moderate, and both forces contribute comparably to the momentum transfer to the particles. Throughout the entrainment process, the pressure-gradient and unsteady drag forces act simultaneously and in the same direction, with comparable magnitudes, jointly contributing to the mechanical work that drives particle uptake. Figure 2C shows the corresponding time evolution of the mechanical work. Although the work is computed from the total force acting on the particles (including pressure-gradient, drag, gravitational, and contact forces), it is dominated during suction by the pressure-gradient and drag contributions; contact forces primarily transmit and redistribute momentum within the granular assembly. Despite the different boundary-velocity profiles in Cases A1 and A2, the accumulated work exhibits very similar trends and final values. This similarity demonstrates that, although the instantaneous force histories differ, the accumulated mechanical work delivered to the particle assembly is nearly the same, consistent with the nearly identical final entrained mass observed in Fig. 1.

This work-based interpretation provides a unified framework for understanding how different boundary kinematics, which may generate very different instantaneous flow and force fields, can nevertheless lead to similar final filling outcomes when the accumulated mechanical work delivered to the particles is comparable. It also clarifies why descriptions based solely on transient pressure fields, as commonly employed in previous studies (e.g., (13)), are insufficient to capture the physical origin of suction-driven entrainment, because they neglect the time-resolved drag, contact forces, and their accumulated mechanical effect. Importantly, the collapse of entrainment behavior onto a work-based measure is not a trivial or self-evident consequence of boundary motion, but emerges despite the strongly unsteady, multiphase, and densely interacting nature of the confined flow. This demonstrates that, even in such complex moving-boundary gas–particle systems, the accumulated mechanical work delivered to the particle assembly along the entrainment direction acts as a robust organizing principle governing particle transport.

**Relationship between boundary displacement and suction-induced entrainment.**

To examine how the extent of boundary motion influences suction-induced particle entrainment, we systematically varied the total downward displacement of the moving bottom boundary while keeping the retreat velocity and the tangential wall motion fixed (Cases B1–B3). Increasing the boundary displacement not only enlarges the void region created beneath the particle bed but also prolongs the time during which boundary-driven unsteady gas flow and the associated fluid–particle interaction forces act in the confined system.

Figure 3A shows five representative snapshots for Case B2, in which the boundary descends beyond the nominal cavity depth (total downward displacement of 3.6 mm), reaches its lowest position at 6.25 ms, and is then raised back to the original cavity height of 3.0 mm. During this extended downward motion, particles are continuously entrained into the enlarged cavity by the suction-induced flow. Figure 3B compares the final particle configurations in the cavity for Cases B1–B3 and clearly shows that cases with extended boundary displacement result in systematically higher final filling amounts. The temporal evolution of the filled mass is shown in Fig. 3C. When the boundary motion exceeds the nominal cavity depth, more particles than the cavity volume can temporarily enter the cavity. As the boundary subsequently returns to the original cavity height, part of this excess material is pushed back to the surrounding container, yet the final retained mass remains larger than in the case without extended displacement. The force and energetic analyses further clarify the origin of this behavior. Figure 3D presents the time histories of the pressure-gradient and drag forces acting on the particles for different boundary displacements. For larger displacements, the duration over which these suction-related forces act is significantly extended, even though their instantaneous magnitudes remain comparable. Figure 3E shows the corresponding time evolution of the mechanical work performed on the particles. The extended boundary motion leads to a clear increase in the accumulated work, demonstrating that a larger total displacement results in a larger mechanical energy transfer from the gas flow to the particle assembly.

Taken together, these results show that increasing the total boundary displacement enhances suction-induced entrainment primarily by prolonging the action time of the pressure-gradient and drag forces, thereby increasing the total mechanical work delivered to the particles. The final filling amount correlates with this accumulated work rather than with the peak values of the instantaneous forces. Thus, a larger boundary excursion produces a longer-lasting unsteady flow, greater integrated force action, and consequently a higher amount of entrained mass, even when the instantaneous suction strength is not increased. This provides a clear physical explanation of why extended boundary motion in confined filling processes leads to enhanced particle uptake through the accumulation of mechanical work over time.

**Conclusions**

In this study, we investigated particle entrainment in confined gas–particle systems driven by moving boundaries using high-fidelity coupled CFD–DEM simulations of unsteady gas flow and discrete particle dynamics. By decomposing the fluid–particle interaction forces during boundary motion, we demonstrated that the so-called suction effect is not solely a consequence of a pressure deficit, as is commonly assumed, but arises from the combined action of pressure-gradient forces and unsteady drag generated by boundary-driven gas acceleration.

Beyond identifying the relevant force components, we showed that the efficiency of particle entrainment is governed by the total mechanical work performed by these forces on the particles. The duration over which pressure-gradient and drag forces act, controlled by the boundary motion, determines the cumulative mechanical work and thus the entrained mass. This work-based interpretation provides a unifying physical framework that reconciles previously phenomenological descriptions of suction-like filling and entrainment processes.

The present results establish a mechanistic link between moving-boundary-induced flow unsteadiness, force accumulation, and particle transport in confined geometries. Such boundary-driven entrainment is ubiquitous in a wide range of systems, from pharmaceutical and nutraceutical powder filling and food processing to chemical engineering operations and natural particulate flows. By clarifying how pressure gradients and drag jointly perform work on particles in these environments, our study offers a general physical basis for predicting and controlling particulate transport in confined gas–particle systems under dynamic boundary conditions.

From a fundamental perspective, we emphasize that in confined gas–particle systems with moving boundaries, the internal flow and force networks are highly unsteady and heterogeneous, involving rapid gas acceleration, strong interparticle contacts, and complex transient structures. Under such conditions, it is far from obvious a priori that particle entrainment can be characterized by a single energetic measure. The present results demonstrate that, despite this apparent complexity, suction-induced entrainment is governed to leading order by the accumulated mechanical work performed on the particle assembly by the total force field along the entrainment direction. This finding explains why descriptions based solely on instantaneous pressure or velocity fields are insufficient, and it establishes mechanical work as a unifying physical quantity for boundary-driven particulate transport in confined geometries. The present work-based scaling is therefore expected to hold in regimes where unsteady, inertia-driven fluid–particle momentum transfer dominates over quasi-static or purely viscous dissipation, i.e., for moderate to large Stokes numbers and confined geometries with rapidly accelerating boundaries.

**Materials and Methods**

We employ an in-house numerical framework (29–31, 33–36) based on a coupled discrete element method (26) and computational fluid dynamics (CFD–DEM) (27) to resolve the unsteady interaction between gas flow and discrete particles under moving boundary conditions. The method incorporates a scalar-field-based representation of solid boundaries using a signed distance function (SDF) (31) together with an immersed boundary method (IBM) (30), which enables consistent treatment of arbitrarily shaped and dynamically moving walls in both the particle and fluid solvers. An implicit time integration scheme is adopted for the interphase drag force to ensure numerical stability in the presence of strong gas acceleration.

The solid phase is described by the DEM (26, 37), in which the translational and rotational motions of each particle are governed by Newton's laws, including contributions from contact forces, gravity, pressure-gradient forces, and fluid drag. Particle–particle and particle–wall contacts are modeled using a linear spring–dashpot formulation with Coulomb friction. The fluid phase is solved using the locally volume-averaged incompressible Navier–Stokes equations. Momentum exchange between the fluid and particles is computed through a drag force closure based on the Ergun (38) and Wen–Yu (39) correlations, with the drag coefficient evaluated as a function of the local void fraction and particle Reynolds number. Moving solid boundaries are represented by the SDF in the DEM and by the IBM in the CFD. In the particle solver, the SDF provides accurate contact detection and force evaluation for arbitrarily shaped moving walls through its local gradient. In the fluid solver, the IBM enforces the no-slip condition on moving boundaries by blending the fluid velocity with the prescribed wall velocity according to the local solid volume fraction, thereby capturing the unsteady flow induced by boundary motion.

**Conditions**

The system consists of a confined domain in which particles are initially contained in an upper compartment. A rigid boundary element forming part of the bottom of the domain retreats downward at a prescribed constant velocity, thereby creating a new void region beneath the particle bed. The motion of this boundary induces an unsteady gas flow in the confined space, which drives the entrainment of particles into the newly formed cavity.

The gas phase is treated as incompressible, and gravity acts uniformly on the particles. The initial particle assembly is at rest and forms a mechanically stable bed prior to boundary motion. The boundary displacement, retreat velocity, and initial filling height are systematically varied to examine their effects on the resulting flow field, fluid–particle forces, and particle entrainment. All remaining physical and numerical parameters are kept fixed unless otherwise stated.

**Acknowledgments**

The authors acknowledge the financial support from the Japan Society for the Promotion of Science KAKENHI (Grant No. 24K22289), and the Social Cooperation Program for Fundamental Technologies on Powder Process Digital Twin in The University of Tokyo. The technical exchanges with Powrex Corporation were beneficial in advancing this research.

38. S. Ergun, Fluid flow through packed columns. *Chem. Eng. Prog.* 48, 89–94 (1952).

39. C. Y. Wen, Y. H. Yu, Mechanics of fludization. *Chem. Eng. Prog. Symp. Ser.* 62, 100–111 (1966).

**Figures and Tables**

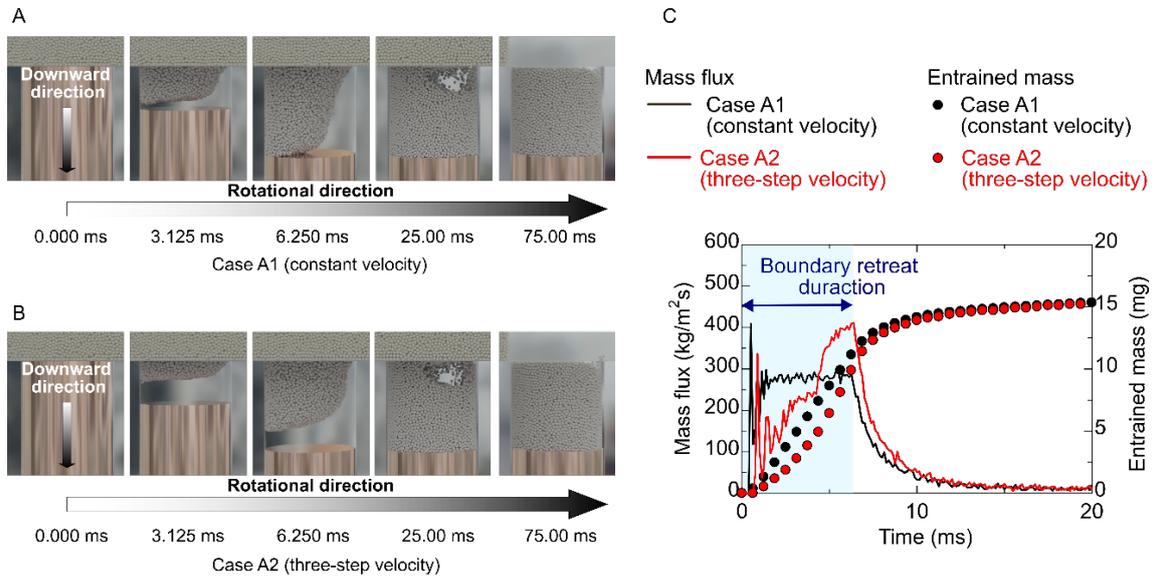

Figure 1. Suction-induced particle entrainment under stepwise constant-velocity boundary motion

(A, B) Particle positions at five representative times for (A) single-stage constant-velocity motion and (B) three-stage piecewise constant-velocity motion of the rigid boundary element. The downward motion of the boundary induces particle entrainment into the cavity.

(C) Particle mass flux and time-integrated entrained mass for Cases A1 and A2. In Case A1, the mass flux remains nearly constant, whereas in Case A2 it increases with each velocity stage. The final entrained mass is nearly identical in both cases.

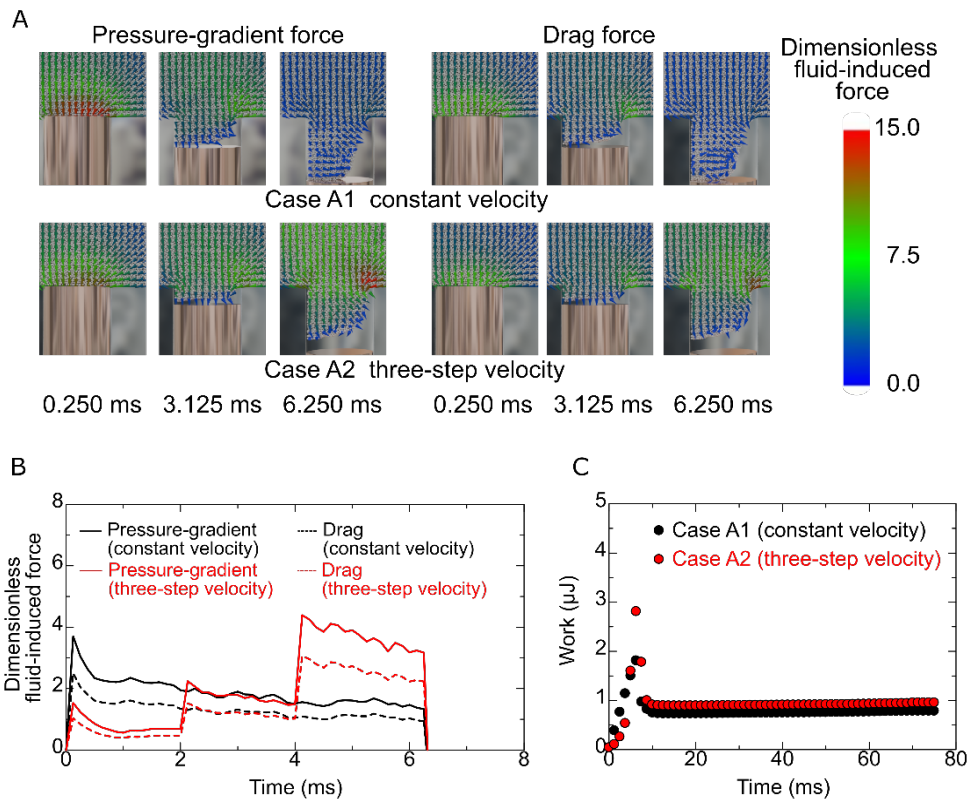

Figure 2. Decomposition of fluid-induced forces and work during suction-driven particle entrainment.

(A) Spatial distributions of the dimensionless pressure-gradient and drag forces acting on particles for Cases A1 and A2.

(B) Components of these forces projected onto the entrainment direction for Cases A1 and A2.

(C) Mechanical work driving particle entrainment (time integral of the power associated with the component of the total force projected onto the entrainment direction).

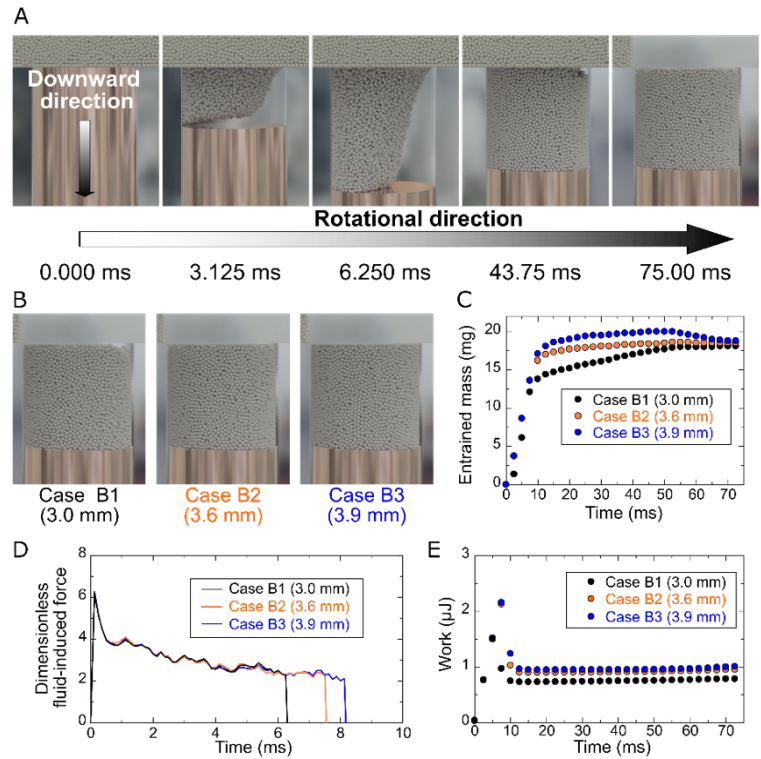

Figure 3. Dependence of suction-induced particle entrainment on boundary displacement.

(A) Representative snapshots of particle distributions during entrainment for an overfilling case (Case B2; total boundary displacement = 3.6 mm).

(B) Final particle configurations in the cavity for Cases B1–B3, illustrating the increase in filling with increasing total boundary displacement.

(C) Temporal evolution of the mass of particles accumulated in the cavity for Cases B1–B3.

(D) Dimensionless pressure-gradient and drag forces along the entrainment direction, showing that larger boundary displacement extends the duration of force application, which contributes to the increase in accumulated work shown in E.

(E) Time evolution of the mechanical work performed on the particles along the entrainment direction, showing that larger boundary displacement leads to greater accumulated work through the prolonged action of pressure-gradient and drag forces, and thus to increased final filling.

21